\documentclass[%
 aip,
 amsmath,amssymb,
 reprint,%
]{revtex4-1}

\usepackage{graphicx}
\usepackage{dcolumn}
\usepackage{bm}
\usepackage{breqn}
\usepackage{hyperref}

\usepackage[utf8]{inputenc}
\usepackage[T1]{fontenc}
\usepackage{mathptmx}

\usepackage{color}

\newcommand{\old}[1]{}

\begin{document}

\preprint{AIP/123-QED}

\title[]{Multi-frequency telecom fibered laser system for potassium laser cooling}

\author{Charbel Cherfan}
\altaffiliation{Universit\'e de Lille, CNRS, UMR 8523 -- PhLAM -- Laboratoire	de Physique des Lasers Atomes et Mol\'ecules, F-59000 Lille, France}
\author{Maxime Denis}
\altaffiliation{Universit\'e de Lille, CNRS, UMR 8523 -- PhLAM -- Laboratoire	de Physique des Lasers Atomes et Mol\'ecules, F-59000 Lille, France}
\author{Denis Bacquet}
\altaffiliation{Universit\'e de Lille, CNRS, UMR 8523 -- PhLAM -- Laboratoire	de Physique des Lasers Atomes et Mol\'ecules, F-59000 Lille, France}
\author{Michel Gamot}
\altaffiliation{Universit\'e de Lille, CNRS, UMR 8523 -- PhLAM -- Laboratoire	de Physique des Lasers Atomes et Mol\'ecules, F-59000 Lille, France}
\author{Samir Zemmouri}
\altaffiliation{Universit\'e de Lille, CNRS, UMR 8523 -- PhLAM -- Laboratoire	de Physique des Lasers Atomes et Mol\'ecules, F-59000 Lille, France}
\author{Isam Manai}
\altaffiliation{Universit\'e de Lille, CNRS, UMR 8523 -- PhLAM -- Laboratoire	de Physique des Lasers Atomes et Mol\'ecules, F-59000 Lille, France}
\author{Jean-Fran\c cois Cl\'ement}
\altaffiliation{Universit\'e de Lille, CNRS, UMR 8523 -- PhLAM -- Laboratoire	de Physique des Lasers Atomes et Mol\'ecules, F-59000 Lille, France}
\author{Jean-Claude Garreau}
\altaffiliation{Universit\'e de Lille, CNRS, UMR 8523 -- PhLAM -- Laboratoire	de Physique des Lasers Atomes et Mol\'ecules, F-59000 Lille, France}
\author{Pascal Szriftgiser}
\altaffiliation{Universit\'e de Lille, CNRS, UMR 8523 -- PhLAM -- Laboratoire	de Physique des Lasers Atomes et Mol\'ecules, F-59000 Lille, France}
\author{Radu Chicireanu}
\altaffiliation{Universit\'e de Lille, CNRS, UMR 8523 -- PhLAM -- Laboratoire	de Physique des Lasers Atomes et Mol\'ecules, F-59000 Lille, France}

\date{\today}

\begin{abstract}
	We describe a compact and versatile multi-frequency laser system for laser-cooling potassium atoms, by frequency doubling a fiber-optic telecom beam ($\simeq 1534$ or $1540$~nm). Low-power fiber-based telecom lasers and components generate a single beam containing the cooling and repumper half frequencies, subsequently amplified by high-power amplifier. A final free-space SHG stage generates a single beam with typically 2.5 W at quasi-resonant frequencies ($\simeq 767$ or $770$~nm) with high-quality mode and ready for laser cooling. This allowed to trap up to $4\times10^9$ $^{41}$K atoms with fast loading times (2.5 s) at sub-Doppler temperatures of 16~$\mu$K. This opens promising perspectives towards versatile and transportable ultracold atom setups.
\end{abstract}

\maketitle

\section{Introduction}\label{sec:level1}

Despite their relative complexity, ultracold atom experiments have become in recent years a widely-used platform for a large range of quantum physics experiments, such as studies of quantum degenerate gases, atomic quantum sensors, and fundamental and applied investigations based on atomic and molecular spectroscopy. A key requirement for that are high power, narrow linewidth laser sources, allowing the capture of large atom numbers in magneto-optical traps (MOT), the common starting point for reaching the quantum degeneracy.

Semiconductor-based laser systems, such as diodes in external cavities and power amplifiers, are cost-effective, but are plagued by limited lifetime and poor spatial mode, while Ti-saphire lasers come at a price. Moreover, in both cases, free-space setups are needed to generate different frequencies for cooling, leading to significant losses.

A fiber-based laser system is thus a good candidate for solving these issues. Several developments~\cite{Stern_2010,Sane_2012,Kwon:20} have been reported in the last decade, and are based on low-power single-frequency lasers which are amplified with powerful telecom C-band erbium-doped fiber amplifiers (EDFA). These techniques are particularly well-suited for cooling potassium or rubidium atoms, with D2 cooling transitions at 767 and 780 nm which can be obtained through second harmonic generation (SHG). However, the amplification and SHG stages are single-frequency, and still require free-space setups for generation of the necessary frequencies. 

In this letter, we demonstrate a flexible multi-frequency laser system architecture for laser cooling, based on telecom fiber amplifier technology. Starting form a low-power single-frequency seeding laser in the telecom domain, the cooling and repumping frequencies are generated simultaneously, using optical fiber components. Cooling and repumper half-frequencies are then combined before optical amplification in a high-power EDFA, followed by a final frequency-doubling stage using a free-space, single-pass nonlinear periodically-poled lithium niobate (PPLN) crystal. Compared to usual setups, our system offers an excellent mechanical stability, and readily contains all the necessary frequencies for laser cooling in a single high-power beam, with an excellent beam quality and spatial mode overlap, and integrates a fast switch-off of the resonant light, without relying upon free-space components.

The performances of this laser architecture is demonstrated by realizing a $^{41}$K potassium isotope magneto-optical trap. This is particularly important due to the narrow excited-state hyperfine structure of the D lines requiring significant repumper laser powers, of the same order of the cooling beams~\cite{Salomon:PRA:14}. Using two similar laser systems, addressing both the D2 (766.701 nm) and D1 (770.108~nm) transitions, we achieve cooling down to sub-Doppler temperatures.

The final system is compact and flexible and, with minor modifications, it could also be used for cooling other potassium isotopes ($^{39}$K and $^{40}$K), as well as for creating degenerate isotope mixtures. This laser design can also be extended for laser cooling of other atomic species (e.g. Rubidium, Mercury, etc.), where fiber lasers/amplifiers have been previously used~\cite{Lienhart2007,Hu:OptExpress:13}. Compact fiber-based laser systems can be used in transportable, possibly space-born experimental applications.

\begin{figure*}[!t]
	\centering
	\includegraphics[width=0.99\linewidth]{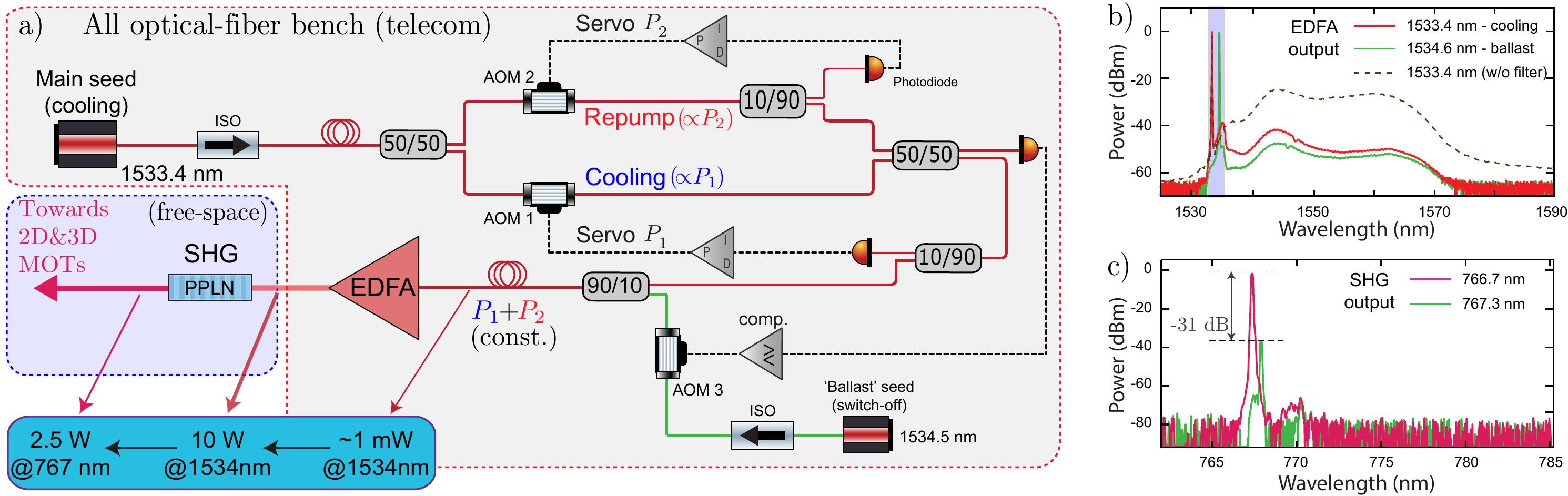}
	\caption{a) Fibered setup for generation of the cooling and repumper frequencies for potassium atoms by SHG (DFB: Distributed Feedback laser ISO: Fiber Optic Isolator; AOM: Acousto-Optic Modulator; PD: Photo-Diode). b) Output spectra of the two-stage EDFA, recorded with an optical spectrum analyzer (OSA), corresponding to the main seed (cooling + repumper, 1533.4 nm) and to the ballast (1534.6 nm). The resolution of the OSA (0.1~nm) does not allow to distinguish the two frequency components (cooling and repumper) of the main seed. The two spectra are recorded in presence of the EDFA mid-stage optical filter (see text), whose range is represented by the shaded area. The dashed line is the output of the EDFA without the filter, and shows a significant increase in ASE. c) Optical spectra of the SHG output. We observe a significant attenuation of the ballast SHG power (767.3 nm) compared to that of the cooling light (766.7 nm), obtained by optimizing the ballast phase mismatch in the PPLN crystal. Together with a $\sim 300$~GHz detuning of the ballast light with respect to the resonance of the $^{41}$K D2 line, this greatly reduces the residual ballast photon scattering rate, and allows us to create a fast and efficient SHG light switch.}
	\label{fig:fig1}
\end{figure*}

\section{\label{sec:level2} Telecom optical fiber setup}

Figure~\ref{fig:fig1} shows a sketch of the optical fiber setup of our system. All the fiber components are polarization-maintaining (PM) and were assembled by fusion splicing, to minimize optical losses. The systems' primary seeding source is a fibered (butterfly) commercial distributed feedback (DFB) laser diode, delivering a power up to 70~mW around 1534~nm. Its frequency is locked by saturated absorption spectroscopy on a ro-vibrational transition of the acetylene molecule (P(15) line at $1534.0991$~nm). This frequency stabilization scheme has been described elsewhere~\cite{Cherfan:20}.

\begin{figure}[!b]
	\centering
	\includegraphics[width=0.8\linewidth]{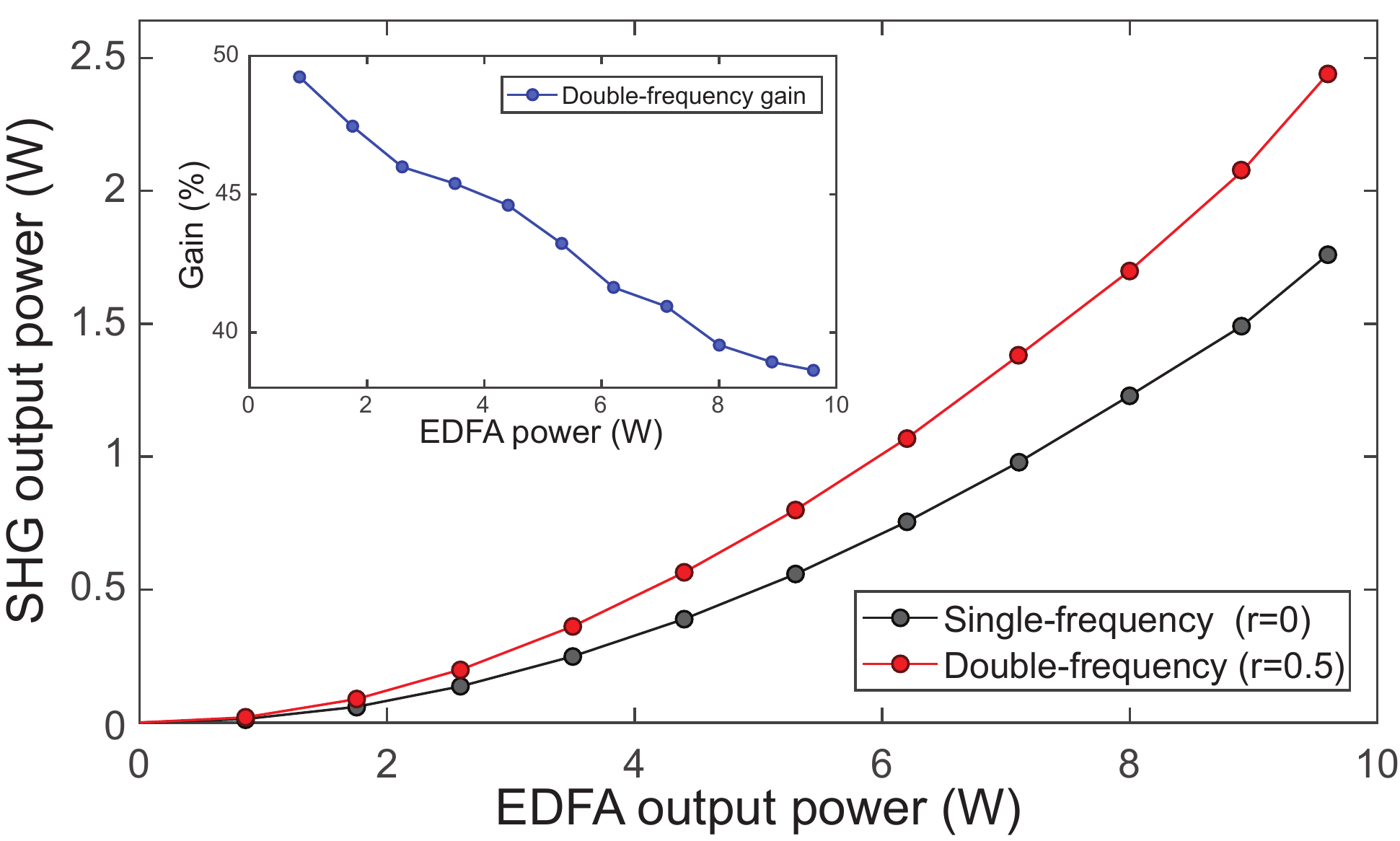}
	\caption{SHG output power as function of the EDFA power, corresponding to the single-frequency seed (blue) and double-frequency seed (red) configurations. Inset: dependence of the relative SHG power increase in the two-frequency seed configuration ($r=0.5$), with respect to the single-frequency case ($r=0$), as a function of the EDFA output power.}
	\label{fig:fig2}
\end{figure}

A fiber-optical isolator protects the master DFB laser against optical feedback. Following the optical isolator, a fiber coupler splits the light in two equal parts, each one being connected to a fibered acousto-optic modulator (AOM). Using different diffraction orders and frequencies allows us to introduce different frequency shifts on each path. The difference between the central frequencies of the two AOMs (-160 MHz and +110 MHz respectively) is chosen close to the ground-state hyperfine splitting of $^{41}$K of $254$~MHz. The two components are then recombined, using a second PM fiber coupler, which generates a two-frequency-component seed for the optical amplifier, corresponding to the cooling and repumping components for magneto-optical trapping.

An important requirement of laser cooling laser systems is the ability to control the frequencies and powers of the cooling and repumper beams. This is achieved in our setup by independently controlling (\textit{via} two separate servo loops) the powers $P_1$ and $P_2$ of the two frequency components which seed the optical amplifier. A first servo loop controls $P_1$, via the RF power which drives AOM 1, and insures a \textit{constant} total power seeding the amplifier, when varying $P_2$ (such that $P_1+P_2={\rm const.}$, typically 1~mW, which insures the saturation regime for the EDFA). The second servo controls $P_2$, and allows to tune the ratio $r\equiv P_2/(P_1+P_2)$ which controls the fraction of repumper light in the final output beam (see below).

\begin{figure*}[t!]
	\centering
	\includegraphics[width=1\linewidth]{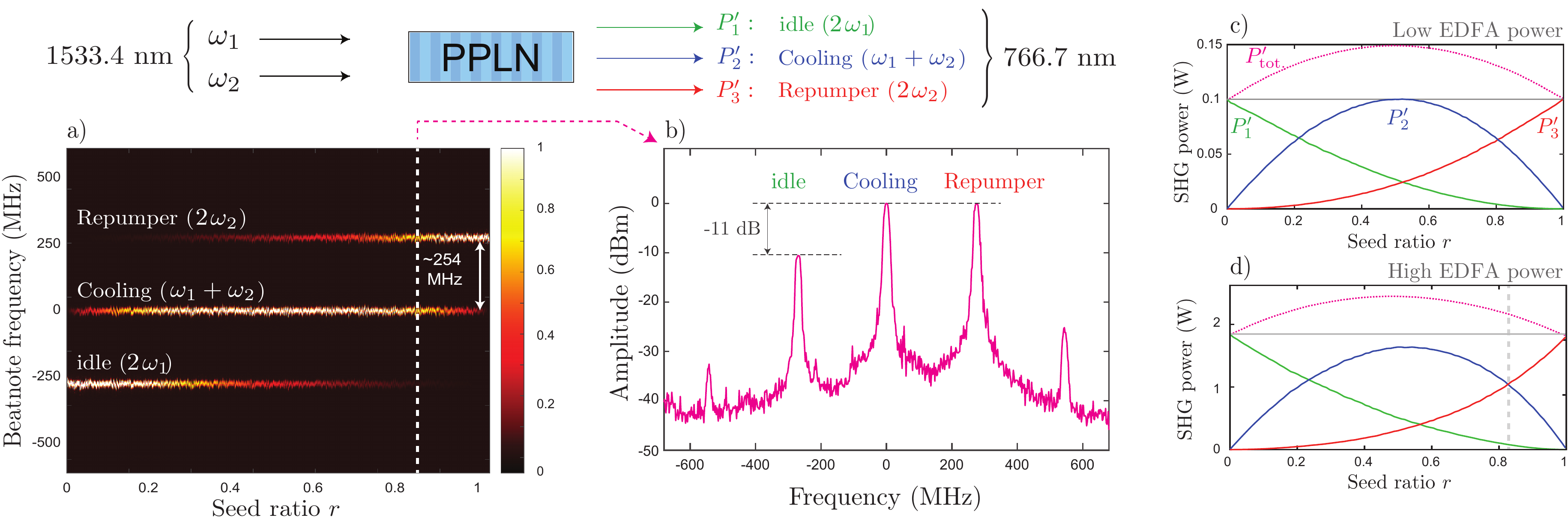}
	\caption{a) Relative power of the SHG spectral components as a function of the input ratio $r$. b) Beatnonte spectrum for $r=0.82$, which corresponds to equal cooling ($\omega_{1}+\omega_{2}$) and repumper ($2\omega_{2}$) powers. (c,d) Dependence of the three main components, $P_1'$, $P_2'$ and $P_3'$, and of the total SHG power $P_{\rm tot.}'$, as a function of the input ratio $r$, at low (c) and high (d) EDFA powers.}
	\label{fig:fig3}
\end{figure*}

The resulting two-component seeding light is used to feed a commercial two-stage EDFA (Quantel, model ELYSA-A-1533-10-P-SN-M-CC), which outputs up to 10 W at 1533.4 nm. This wavelength stands on the lower edge of the gain region of the ytterbium-doped amplification fiber (see Fig.~\ref{fig:fig1}.b), which somewhat limits the suppression of the amplified spontaneous emission (ASE). To allow efficient EDFA output at this wavelength, a narrow mid-stage optical filter (2 nm FWHM, centered at 1534 nm) was added to suppresses the ASE before the final amplification stage (Fig.~\ref{fig:fig1}.b). The output beam of the EDFA is directly focused to $w_0\simeq 80\ \mu$m ($1/e^2$ diameter) inside a 50~mm long PPLN crystal, in a single pass configuration. The SHG phase-matching condition at the EDFA wavelength is obtained by temperature-tuning the crystal to $\sim 119 \ ^{\circ}$C. The $766.7$~nm SHG output light readily contains the necessary frequencies for laser cooling, and is further split in several parts, and directed towards the cold-atom part of our setup \textit{via} polarization-maintaining optical fibers. This simple configuration represents the only free-space part of our system. As a consequence, our system presents a remarkable mechanical stability, compared to usual free-space laser setups -- e.g. no realignment of fiber-launch system being necessary over long time periods.

Our setup provides an efficient and convenient way to realize fast switch-off of the cooling and repumper, which is particularly important after of the laser cooling sequence. The can be achieved by rapidly turning off the main seeding light (AOMs 1 and 2). However, operating the EDFA in absence of seeding light may induce permanent damage to it. Thus, we utilize a secondary seed laser (`ballast'), see Fig.~\ref{fig:fig1}, whose presence is controlled by a fast comparator circuit and an AOM switch, which injects the ballast light towards the EDFA whenever both the AOMs 1 and 2 are switched off. The `ballast' laser parameters were carefully chosen to minimize its effect on the laser cooled atoms, by tuning its wavelength to a value which maximizes the phase mismatch in the PPLN crystal ($1534.6$~nm), thus reducing the SHG efficiency to a level of $\sim 10^{-3}$ (Fig.~\ref{fig:fig1}.c). Compared to free-space AOM switches, commonly used in cold atom setups, this technique has the advantage of avoiding unnecessary power losses of the cooling light at the output of the SHG crystal.

\section{\label{sec:level3}Spectral characterization of the laser output}

Throughout laser cooling experimental sequences, it is important to control precisely the power levels of the resonant light components, i.e. the cooling and repumper light, which in our case correspond to the multi-frequency output of the nonlinear SHG stage. We shall now focus on the spectral characterization of the SHG output as a function of the frequency composition used at the EDFA input -- with corresponding angular frequencies (powers) $\omega_1$ and $\omega_2$ (resp. $P_1$ and $P_2$) -- and compare our results to a simple theoretical model.

\subsection{Second harmonics and sum-frequency generation}

Figure~\ref{fig:fig2} (red curve) shows the dependence of the \textit{total} SHG power (767 nm) as a function of the EDFA output power (1534 nm), in two configurations. For a single-frequency seed configuration ($100~\%$ of either $P_1$ or $P_2$) we obtain an approximately quadratic dependence of the SHG power, expected in absence of pump depletion effects, as a function of the pump power, corresponding to a maximum value of 1.7~W for 10~W at the EDFA output. Intristingly, while the EDFA power remains constant, the SHG output power increases noticeably when two frequencies are used to seed the optical amplifier. In this configuration, the maximum total SHG power is obtained when the two seed frequency components are used in equal parts ($P_1\simeq P_2$, see Fig.~\ref{fig:fig2}, black). The inset of Fig.~\ref{fig:fig2} shows the relative SHG gain in the double-frequency configuration, with respect to the single-frequency configuration, as a function of the EDFA power. At low EDFA power, the relative multi-frequency SHG power increase is $\simeq 50\%$. For higher IR powers the multi-frequency gain follows a slowly-decreasing trend, down to $\simeq 35\%$ at maximum EDFA power.

To investigate this behavior, we performed a careful spectral characterization of the different stages of the system. First, we checked that the EDFA input seed ratio $r$ is preserved at its output, and its total power remains constant and independent of $r$. This behavior is somewhat expected, because the optical amplifier works in the saturation regime with respect to the seed power, and its gain is almost identical for $\omega_1$ and $\omega_2$, for small values of $\Delta \omega$. This shows that the multi-frequency power increase is attributed to the SHG stage.

To analyze the spectral composition after the SHG, we used an optical beat note signal obtained by mixing a small fraction of the output of the nonlinear PPLN crystal with an independent $767$~nm (single-frequency) reference laser. The frequency difference between the two lasers was set to $\sim 1$~GHz, which allows to observe all the spectral components of our laser. The beat note is detected using a fast ($2$~GHz bandwidth) photodiode connected to an AC-coupled electrical spectrum analyzer, for different values of $r$ (see Fig.~\ref{fig:fig3}.a). We observe three main SHG output components, separated by $\Delta \omega$, which correspond to the main nonlinear mixing processes expected to occur in the PPLN crystal, i.e. frequency-doubling of $\omega_1$ and $\omega_2$, as well as sum-frequency generation $\omega_1+\omega_2$. Fig.~\ref{fig:fig3}.b shows an example of the measured beat note signal, corresponding to a ratio $P_2/(P_1+P_2)\simeq 0.82$, where two of the SHG components are dominant and occur in equal proportions. This configuration is particularly interesting in our setup, and is very close to the optimal laser cooling parameters for potassium~\cite{SupplMat}.

\subsection{Model}

The spectral behavior of our laser system can be understood using a simple model. Let us consider the electrical fields of the two EDFA output components, in complex representation, before SHG:
\begin{equation}
\begin{cases}
E_{1}=\mathcal{E}_{0}\sqrt{1-r}e^{i\omega_{1}t}+{\rm cc.}\\
E_{2}=\mathcal{E}_{0}\sqrt{r}e^{i\omega_{2}t}+{\rm cc.}
\end{cases}, \ \ E_{tot}=E_{1}+E_{2}
\end{equation}
where $r$ (defined above) is the power ratio between the $E_2$ input component to the total input power $P_0$, corresponding to $E_{tot}$. At the output of the $\chi^{(2)}$ crystal (PPLN), the electric field generates electrical field components $E'_{i,j}=\chi^{(2)}E_{i}E_{j}$ ($i,j=1,2$), such that the total electric field can be written as: 
\begin{equation}
E'_{tot}\propto\chi^{(2)}E_{tot}^{2}=\chi^{(2)}\left(\mathcal{E}_{0}\sqrt{1-r}e^{i\omega_{1}t}+\mathcal{E}_{0}\sqrt{r}e^{i\omega_{2}t}+{\rm cc.}\right)^{2}
\end{equation}
Developping the last term, and neglecting the DC and low-frequency  ($\omega_{1}-\omega_{2}$) components, we easily obtain:
\begin{dmath}
	E'_{tot}\equiv E'_{1}+E'_{2}+E'_{3} \propto\chi^{(2)}\mathcal{E}_{0}^{2}\left[(1-r)e^{i2\omega_{1}t}+2\sqrt{r(1-r)}e^{i(\omega_{1}+\omega_{2})t}+re^{i2\omega_{2}t}+{\rm cc.}\right]
\end{dmath}

Using this expression, we obtain that the total power at the output of the SHG is given by:
\begin{equation}
P'_{tot}(r)\equiv P'_1+P'_2+P'_3\propto P_{0}^2 \left[(1-r)^{2}+4r(1-r)+r^{2}\right]\label{eq:eq4}
\end{equation}

The frequency-doubling components, $P'_1$ and $P'_3$, have a quadratic dependence on the power of the respective SHG input components: $P'_1\propto P_0^2 (1-r)^2$ and $P'_3\propto P_0^2 r^2$, respectively. The sum-frequency component $P'_2$ is proportional to the product of the SHG input power components $P'_2\propto P_{0}^2\times 4r(1-r)$, and has a maximum value for $r=0.5$, i.e. for an even distribution between the SHG inputs.

Eq.\eqref{eq:eq4} is in excellent agreement with our measurements performed at low values of SHG input power ($\sim 500$ mW), shown in Fig.~\ref{fig:fig3}.c. At higher output powers, we observe deviations from eq.\eqref{eq:eq4}, as shown in Fig.~\ref{fig:fig3}.d. While the maximum SHG power is also reached for $r\simeq 0.5$, the multi-frequency SHG gain is somewhat lower (see also Fig.~\ref{fig:fig2}). This is probably due to a phase mismatch caused by thermal effects in the PPLN crystal~\cite{Kang:ApplOpt:17} and, to a lower extent, to pump depletion effects. More importantly, for EDFA input power ratios $r\simeq 0.8$, approximately $96\%$ of the total SHG power is distributed equally between the two spectral components of interest (cooling, $P'_2$, and repumper, $P'_3$). This configuration is close to the one typically used in potassium MOTs. The remaining $\approx 4\%$ of the power is contained in the other harmonics, which are off-resonance with respect to the cooling/repumper transitions, and are not expected to contribute to the cooling process.

\section{Laser-cooling $^{41}$K to sub-Doppler temperatures}

We tested the performance of our laser architecture in a $^{41}$K experiment, with the goal of achieving suitable starting conditions for evaporative cooling towards Bose-Einstein condensation. The $^{41}$K laser cooling phase is performed in two main steps: the first step uses a laser system tuned to the potassium D2 line ($766.7$~nm) to load a 3D magneto-optical trap fed by a slow beam of atoms issued by a 2D MOT. After a short compressed MOT stage, we then utilize a second laser system, built to address the D1 line at $770$~nm, to further cool the atoms in a gray molasses stage. The experimental setup and cooling sequences are described in greater detail in the Supplementary materials~\cite{SupplMat}.

\begin{figure}[!t]
	\centering
	\includegraphics[width=0.75\linewidth]{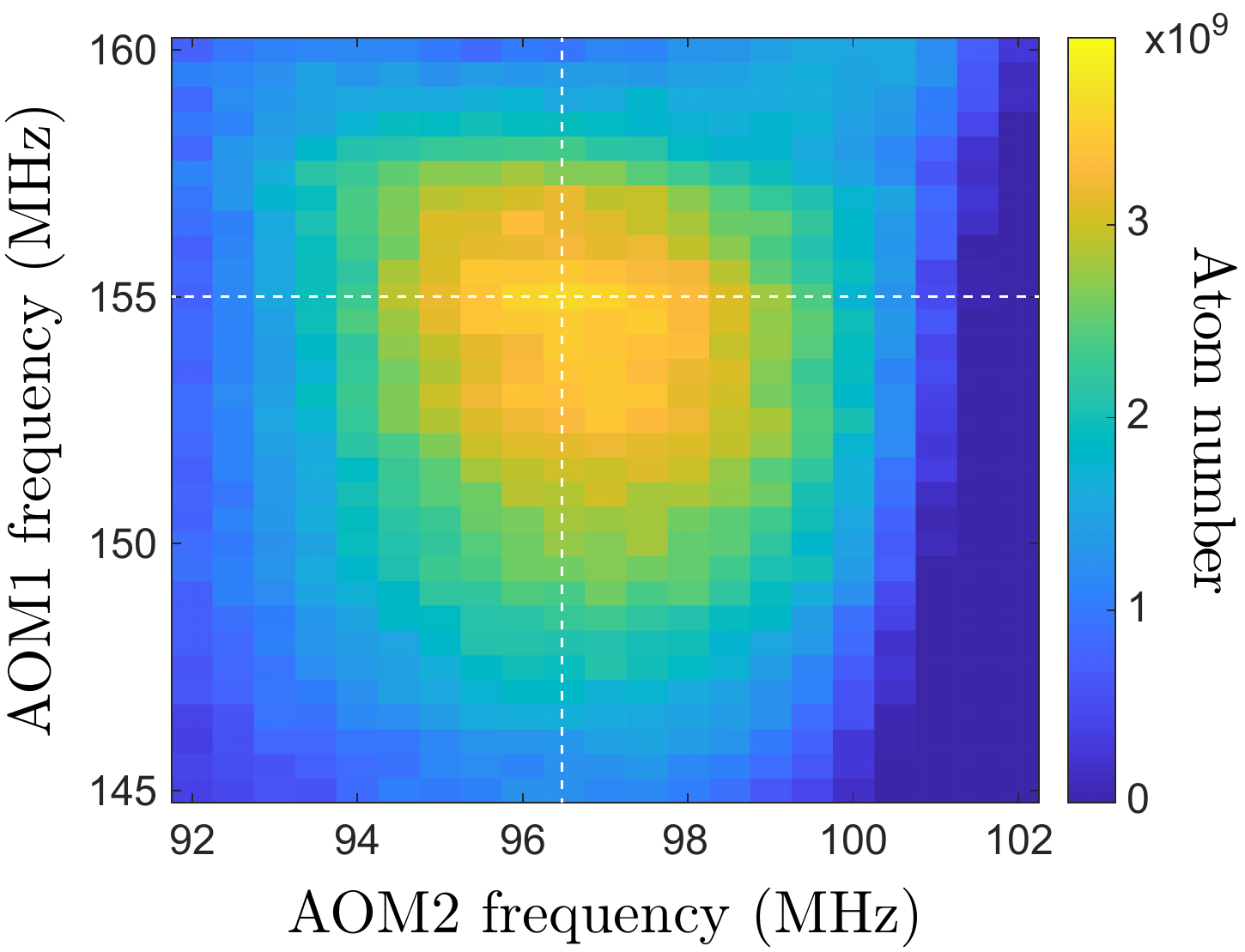}
	\caption{Dependence of the MOT atom number as a function of the two seeds' AOM frequencies ($\omega_1$ and $\omega_2$), which control detunings of the cooling and repumper components. The maximum atom number is $N\simeq 4.10^{9}$ atoms, and was obtained with a cooling/repumper power ratio of $45\%$, and for corresponding detunings of $-33$~MHz and $-25$~MHz respectively~\cite{SupplMat}.}
	\label{fig:fig4}
\end{figure}

Fig.~\ref{fig:fig4} shows a color plot representing the number of trapped atoms as a function of the two AOM frequencies controlling the respective detunings of the cooling and repumper components. The maximum number of atoms is achieved for $\omega_1/2\pi=155$~MHz and $\omega_2/2\pi=96.5$~MHz, which correspond to detunings for the cooling (resp. repumper) light of $-5.5$ (resp. $-4.2$) in units of natural linewidth of the potassium D2 transition ($\Gamma/2\pi=6.036$~MHz). A relatively weak magnetic field gradient ($5.5$~G/cm along the axial direction of the coils) is used, to keep a relatively low atomic density and reduce light-assisted collisions processes, which limit the atoms number in potassium MOTs~\cite{Prevedelli:PRA:99}. While further lowering the magnetic field gradient can lead to up to a two-fold increase in atom number, this also significantly increases the MOT loading time.

At this stage on the experiment, the atomic cloud temperature is 2~mK. This value is significantly higher than the K Doppler limit (144~$\mu$K), because of large depumping/heating rates induced by the small hyperfine structure of the excited state $^4$P$_{3/2}$ of the potassium D2 line~\cite{Salomon:PRA:14}. As previously demonstrated for the $^{39}$K isotope~\cite{Salomon:PRA:14}, further laser cooling can be achieved using the 770~nm D1 transition, using of the so-called `gray-molasses' to cool the atoms through optical pumping to motional hyperfine dark states, corresponding to the zero-velocity class~\cite{RioFernandes:EPL:12,Nath:PRA:13,Salomon:EPL:13}. We implemented this technique using a second multi-frequency laser based on the same architecture, working at $\sim 770$~nm. The atom transfer and the gray molasses, described in the supplementary material, allowed us to reduce the cloud temperature down to 16~$\mu$K, with a phase-space density of $7\times 10^{-5}$ (see~\cite{SupplMat}). These represent promising starting conditions for performing evaporative cooling to achieve Bose-Einstein condensation of $^{41}$K atoms.

\section{Conclusions and outlook}

In this letter, we demonstrated an improved multi-frequency, high-power laser setup applicable for laser cooling. The power levels achieved with this setup are sensibly higher than what has typically been reported in other potassium BEC experiments~\cite{Kishimoto:PRA:09,Salomon:PRA:14,Chen:PRA:16}. Our telecom fiber-based laser system provides several advantages compared to traditional solid-state setups -- such as mechanical robustness, low maintenance, and excellent output beam quality. We demonstrated its applicability in cold-atom experiments, by implementing two laser-cooling stages using both D transitions of potassium.

Our setup should be easily extended, with minor changes, to the other isotopes of potassium, or to other atomic species such as Rubidium and Mercury. Several improvements may be conceivable in the future. Higher SHG efficiencies may be achieved using more powerful fiber amplifiers as well as a SHG enhancement cavity, as has been recently reported in the case of single-frequency laser system~\cite{Kwon:20}. Another interesting direction could be further explored towards a fully fiber-integrated setup, by using fiber-coupled waveguides for the SHG stage~\cite{Nishikawa:09}. This could open new avenues towards transportable cold-atoms setups, with possible metrological or fundamental-science applications.

\section*{Supplementary material}
See supplementary material for a detailed description of the cold atom setup and different experimental stages used for sub-Doppler laser cooling $^{41}$K atoms.

\section*{Acknowledgments}
This work was financially supported by Agence Nationale de la Recherche through Research Grant MANYLOK No. ANR-18-CE30-0017, the Labex CEMPI (Grant No. ANR-11-LABX-0007-01), the I-SITE ULNE (ANR-16-IDEX-0004 ULNE - QUITOPS project), the Programme Investissements d'Avenir (ANR-11-IDEX-0002-02, reference ANR-10-LABX-0037-NEXT), the Ministry of Higher Education and Research, Hauts-de-France Council and European Regional Development Fund (ERDF) through the Contrat de Projets Etat-Region (CPER Photonics for Society, P4S).


%

\end{document}


\preprint{AIP/123-QED}

\title[]{Supplementary Material: Multi-frequency telecom fibered laser system for potassium laser cooling}

\author{Charbel Cherfan}
\altaffiliation{Universit\'e de Lille, CNRS, UMR 8523 -- PhLAM -- Laboratoire	de Physique des Lasers Atomes et Mol\'ecules, F-59000 Lille, France}
\author{Maxime Denis}
\altaffiliation{Universit\'e de Lille, CNRS, UMR 8523 -- PhLAM -- Laboratoire	de Physique des Lasers Atomes et Mol\'ecules, F-59000 Lille, France}
\author{Denis Bacquet}
\altaffiliation{Universit\'e de Lille, CNRS, UMR 8523 -- PhLAM -- Laboratoire	de Physique des Lasers Atomes et Mol\'ecules, F-59000 Lille, France}
\author{Michel Gamot}
\altaffiliation{Universit\'e de Lille, CNRS, UMR 8523 -- PhLAM -- Laboratoire	de Physique des Lasers Atomes et Mol\'ecules, F-59000 Lille, France}
\author{Samir Zemmouri}
\altaffiliation{Universit\'e de Lille, CNRS, UMR 8523 -- PhLAM -- Laboratoire	de Physique des Lasers Atomes et Mol\'ecules, F-59000 Lille, France}
\author{Isam Manai}
\altaffiliation{Universit\'e de Lille, CNRS, UMR 8523 -- PhLAM -- Laboratoire	de Physique des Lasers Atomes et Mol\'ecules, F-59000 Lille, France}
\author{Jean-Fran\c cois Cl\'ement}
\altaffiliation{Universit\'e de Lille, CNRS, UMR 8523 -- PhLAM -- Laboratoire	de Physique des Lasers Atomes et Mol\'ecules, F-59000 Lille, France}
\author{Jean-Claude Garreau}
\altaffiliation{Universit\'e de Lille, CNRS, UMR 8523 -- PhLAM -- Laboratoire	de Physique des Lasers Atomes et Mol\'ecules, F-59000 Lille, France}
\author{Pascal Szriftgiser}
\altaffiliation{Universit\'e de Lille, CNRS, UMR 8523 -- PhLAM -- Laboratoire	de Physique des Lasers Atomes et Mol\'ecules, F-59000 Lille, France}
\author{Radu Chicireanu}
\altaffiliation{Universit\'e de Lille, CNRS, UMR 8523 -- PhLAM -- Laboratoire	de Physique des Lasers Atomes et Mol\'ecules, F-59000 Lille, France}

\date{\today}
\maketitle

This Supplementary Material describes in greater detail the experimental apparatus used for laser cooling and trapping $^{41}$K atoms down to sub-Doppler temperatures. The setup uses a `standard' 3D MOT fed by a slow atomic beam produced by a 2D MOT.

The vacuum system consists of two chambers: the 2D-MOT chamber, filled with a low-pressure potassium vapor, and an ultra-high vacuum science chamber for ultracold-gas experiments. The 2D MOT is a commercially-available system, developed by the LNE-SYRTE laboratory in Paris, France. The vacuum chamber is machined in a titanium block and is fitted with indium-sealed AR-coated view ports. The atomic vapor is created  using a potassium ampoule heated to $80^\circ$C and connected to the 2D MOT chamber through a CF16 valve. A temperature gradient helps the potassium migrate from the ampoule to the 2D MOT chamber, which is maintained at a lower temperature ($50^\circ$C) to prevent the potassium to stick to the vacuum viewports. The 2D MOT chamber is pumped by a small 2l/s ion pump, which maintains a pressure of $5\times10^{-8}$~mBar. The 3D MOT is created in the second vacuum chamber pumped by two getter-ion pumps (NexTorr, models D500 and D300) and connected to the 2D MOT \textit{via} a differential pumping stage. The pressure is on the order of a few $10^{-11}$~mBar, with a MOT lifetime >30 seconds. The distance between the 2D MOT output and the 3D MOT chamber is $26$~cm.

The $^{41}$K laser cooling sequence is performed in three stages, presented below. For that, we utilize two multi-frequency laser systems built using the architecture presented in this Letter. The first laser system, described in the main paper, addresses the D2 transition at $766.7$~nm, and is mainly used for operating the `standard' 2D and 3D MOTs. The second laser system, seeded by a $1540.2$~nm laser diode, addresses the potassium D1 transition at $770.1$~nm for subsequent cooling stages. The performance of the two systems are quite similar, with typically $\sim2.6$~W available at the output of the D1 laser.

\begin{table*}
	\begin{tabular}{c c c ccccccccc} 
		\hline\hline 
		& $B'$ (G/cm) \  &\   $\delta^{c}_{D2}/\Gamma$ \ & $\delta^{r}_{D2}/\Gamma$ \ & $\delta^{c}_{D1}/\Gamma$ \ & $\delta^{r}_{D1}/\Gamma$ \ &  $P_{D2}$ (mW) \ &  $P_{D1}$ (mW) \ & $R_{D2} (\%)$ \ & $R_{D1} (\%)$ \ & N $ \times 10^{9}$ \ & T ($\mu$K)
		\\ 
		\hline 
		\\
		3D MOT (D2) & 5.5	& -5.5 & -4.2 & - & - & 240 & - & 45 &- &  4  & 2000 \\
		
		`Hybrid' CMOT & 5.5 $\rightarrow$ 25& -	&-5.13 & -1.8 & - & 60  & 270 &100 & 0 & 3.7   & 530 \\
		
		Gray molasses (D1) & -& - & - & 1.85 & 1.85 & - & 270 $\rightarrow$24 & - & 15 &3 &16 \\
		\hline 
	\end{tabular}

	\caption{Optimal parameters during the different laser cooling stages for $^{41}$K. $B'$ is the magnetic field gradient, $\delta$ the frequency detuning , $\Gamma$ the natural linewidth of the transition ($\Gamma/ 2\pi=6.036$~MHz for the D2 transition and $\Gamma / 2\pi=5.95$~MHz for the D1 transition), $P$ is the total laser power of the beams sent into the chamber, $R$ the power ratio between the cooling and the repumper frequency components. Superscripts $c$ and $r$ refer to the cooling and repumper respectively.} 
	
	\label{tab:LPer}
\end{table*}

\section{\label{sec:level1} 2D $\&$ 3D magneto-optical traps}

\begin{figure}
	\centering
	\includegraphics[width=0.9\linewidth]{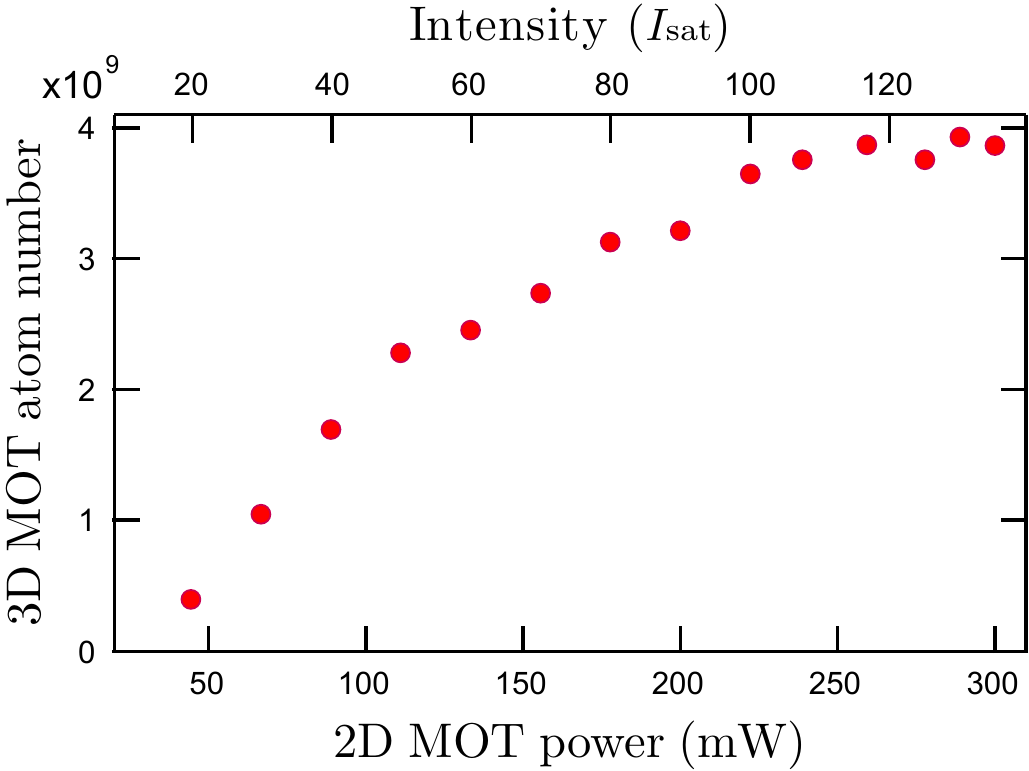}
	\caption{Atom number loaded in the 3D MOT as a function of the total power of the 2D MOT beams. We observe a saturation which occurs above $\sim 250$~mW  (i.e. a maximum light intensity of $\sim 120\times I_{\rm sat}$, with $I_{\rm sat}=1.75$~mW/cm$^2$ the saturation intensity of the D2 line).}
	\label{fig:Fig1SM}
\end{figure}

The MOT loading phase utilizes only the D2 laser system, whose output is split in two main parts: approximately $55 \%$ of the total power is used to generate the 2D MOT beams, while the remaining $45\%$ is used for the 3D MOT. The 2D MOT beam is injected in a PM optical fiber which is fusion-spliced to a $50-50 \%$ PM fiber coupler. This provides independent cooling beams for the two orthogonal directions of the 2D MOT. Each output of the coupler is collimated using a cylindrical telescope (aspect ratio 1:3), and separated into three parts, using polarization beam splitters and half waveplates, for a total interaction length of $\simeq 10$~mm. All 2D MOT optics are directly mounted onto the vacuum chamber, and each beam is retro-reflected after a first passage through the chamber. Two pairs of rectangular coils create a magnetic field gradient for the 2D MOT, with an optimal value of $12$~G/cm. A low-power ($1$ mW) push beam, blue detuned from the cooling transition by 3.5~$\Gamma$, helps enhance the flux of atoms directed towards the 3D MOT by a factor of $\simeq 2$. Fig.~\ref{fig:Fig1SM} shows the dependence of the total atom number captured by the 3D MOT, as a function of the total power used for the 2D MOT beams.

Our 3D MOT design uses independent laser beams obtained by splitting the dedicated laser power in three equal components, using free-space polarizing beamsplitters and half waveplates. Each component is then injected in PM optical fibers and directed towards the science chamber. The output of each fiber is separated in two, using fusion spliced to $50-50 \%$ PM fiber couplers, which creates corresponding pairs of counter-propagating beams in each space direction. The MOT beams are collimated to a diameter ($1/e^2$) of $11.5$~mm, yielding a total six-beam intensity of $29\times I/I_{\rm sat}$, with $I_{\rm sat}=1.75$~mW/cm$^2$ the saturation intensity of the potassium D2 transition. The 3D MOT magnetic field gradient is created by a pair of water-cooled coils is anti-Helmholtz configuration. We utilize a relatively low value of magnetic field gradient (typically $5.5$~G/cm), which helps us to reduce light-assisted inelastic collisions, which are known to limit the atoms number in potassium MOTs~\cite{PrevedelliPRA1999}. Thanks to this configuration, we are able to load $2.5 \times 10^{9} $ atoms with a temperature of $2$~mK and typical $1/e$ loading times of $2.5$ to $3.5$~s.

Interestingly, the number of MOT atoms can be further increased by shining a retro-reflected, low-power ($\simeq1$~mW) resonant beam of D1 light in the transverse direction of the 2D MOT atomic beam (in practice, this beam is also used in later stages of the experiment, for optically pumping the atoms in the $F=2, m_F=+2$ state, prior to the loading atoms in a magnetic quadrupole trap -- not discussed here). This effect is likely related to a decrease in the MOT density, which further lowers the light-assisted collisional loss rate, leading to an increase by $60\%$ the number of atoms in the MOT. In presence of this additional beam, the number of atoms in the MOT reaches $4 \times 10^{9}$, with a phase space density of $8\times 10^{-9}$.

\section{\label{sec:level2} Hybrid compressed-MOT}
Following the MOT loading stage, we perform a transient compressed MOT (C-MOT) phase which helps increase the density of the $^{41}$K cloud. This is achieved by increasing the magnetic field gradient during a few milliseconds. We opted for a `hybrid' C-MOT configuration, which uses a combination of blue-detuned D1 cooling light ($770.108$~nm) close to the $F=2 \to F'=2$ transition, and D2 repumper ($F=1 \to F'=2$). This scheme has been previously reported in the case of $^{39}$K atoms~\cite{Salomon:PRA:14}, and was interpreted as a combination between a gray-molasses–type sub-Doppler cooling~\cite{Fernandes:EPL:2012,Nath:PRA:13} occurring from the blue-detuned D1 light, and a magneto-optical trapping force generated by the magnetic field gradient and the red-detuned D2 light~\cite{Salomon:PRA:14}. Compared to a `standard' C-MOT performed directly on the D2 transition, the hybrid C-MOT configuration allows us to further reduce the temperature of the cloud, which leads to a significant increase of the phase-space density with respect to the 3D MOT (see below).

\mbox{}

\begin{figure} [!h]
	\centering
	\includegraphics[width=1\linewidth]{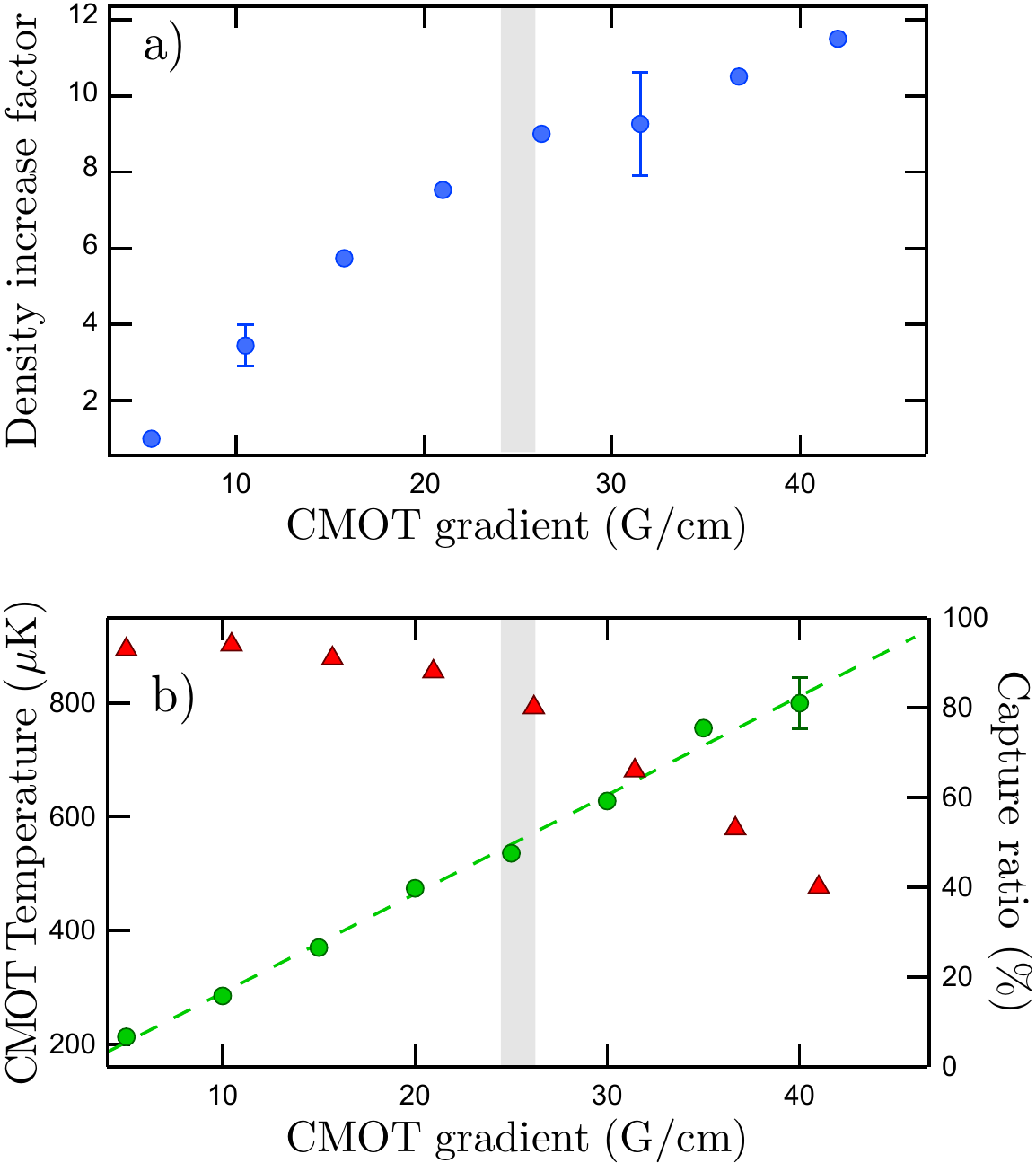}
	\caption{a) Density increase factor at the end of the CMOT phase as a function of the final CMOT magnetic field gradient. b) Temperature (green circles) as a function of the magnetic field gradient $B'$. The figure also shows (red triangles) the dependence on $B'$ of the atom capture efficiency, following the subsequent cooling stage (gray molasses -- see next section), which drops rapidly beyond a certain threshold value ($B'\simeq 25$~G/cm, represented by the shaded area).}
	\label{fig:Fig2SM}
\end{figure}

The optimal parameters for the hybrid CMOT are shown in Table\ref{tab:LPer}. We use both the D1 and D2 lasers in single-frequency configurations: the cooling D1 laser is used at full power, whereas the D2 repumper is set to $~20\%$ of the total available power. Subsequently, the power of the D2 laser is linearly ramped down to zero, during $10$~ms.

Figure~\ref{fig:Fig2SM}.a) shows the dependence of the atomic density increase factor, after the CMOT phase, on the magnetic field gradient $B'$. We observe more than a tenfold increase in density upon the variation of the magnetic field gradient $B'$. The CMOT induces relatively small losses in atom number ($\lesssim 10 \%$), and its final density does not depend on the density of the 3D MOT. However, increasing $B'$ also leads to an approximately linear increase of the CMOT temperature (Fig.\ref{fig:Fig2SM}.b). This temperature increase turns out to be detrimental, above a certain value, to the transfer of the CMOT atoms into the subsequent optical molasses (see next section for details). For this reason, we chose to operate the CMOT at $B'=25$~G/cm, represented by the shaded areas in Fig.\ref{fig:Fig2SM}. In these conditions, we obtain a final CMOT temperature of $500\ \mu$K and a nine-fold density increase with respect to the initial MOT (i.e. a $\times 70$ increase in phase-space density).

\section{\label{sec:level3} Gray molasses}

To further cool the potassium atomic cloud below the Doppler limit, we implement a gray-molasses stage using the potassium D1 transition. Although quite general, this technique has been particularly successful to sub-Doppler cool lithium and potassium bosonic isotopes, where the narrow D2 excited state hyperfine structure severely affects the efficiency of `standard' optical molasses~\cite{Nath:PRA:13}. The gray molasses technique circumvents this limitation, by combining two effects~\cite{Salomon:EPL:13}: at first, the atoms with non-zero velocity are effectively slowed down through Sysiphus cooling mechanisms~\cite{Dalibard:JOSAB:89}, until approaching $v\simeq0$, at which point they are optically pumped to a dark state, which effectively decouples them from the near-resonant cooling light (similar to the velocity-selective population trapping mechanisms~\cite{Aspect:PRL:1988}).

\begin{figure}[!h]
	\centering
	\includegraphics[width=0.65\linewidth]{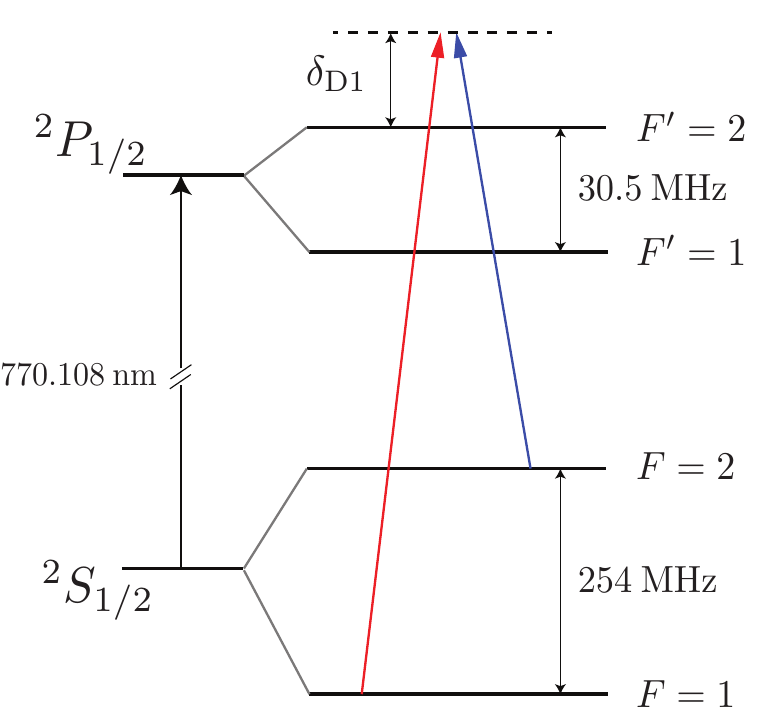}
	\caption{$^{41}{\rm K}$ hyperfine levels of the D1 line (not to scale), and the two light fields corresponding to the gray molasses configuration, addressing the D1 cooling (blue) and repumper (red) transitions.}
	\label{fig:Fig3SMa}
\end{figure}

In our case, the gray molasses is formed by two light fields in a $\Lambda$-configuration (see Fig.~\ref{fig:Fig3SMa}), from the two hyperfine ground states $F=1$ and $F=2$ to the blue side of the $F'=2$ excited state of the D1 line. When the Raman condition for the two light fields is satisfied (i.e. their frequency difference equals the ground state hyperfine splitting), the atoms with near-zero velocity can be optically pumped in a `hyperfine dark state', a coherent superposition of the two hyperfine ground states $F=1$ and $F=2$ no longer coupled with the light field. This process efficiently populates the $v=0$ velocity class, while limiting heating through spontaneous emission of these atoms, thus leading to a dramatic reduction in temperature of the cloud. More details on the hyperfine gray molasses mechanisms can be found in~\cite{Nath:PRA:13}.

\mbox{}

In our experiment, during the gray molasses phase the D2 laser and magnetic field gradient are switched off, and the D1 beam is switched on in a two-frequency configuration, with both cooling and repumper frequencies present. The temperature and density of the gray molasses depend crucially on the fulfillment of the Raman detuning condition of the hyperfine levels of the ground state with respect to the excited state. We thus fix the frequency difference between the D1 cooling and repumper to $\Delta \omega=254$~MHz, with the possibility of scanning the overall detuning of both beams with respect to the excited $^2 P_{1/2}$, $F'=2$ state. Optimal results are obtained when both cooling and repumper beams are blue detuned by $\delta_{\rm D_1}=1.85$~$\Gamma$, and the cooling/repumper power ratio equals $\sim 15 \%$.

\begin{figure}[!t]
	\centering
	\includegraphics[width=1\linewidth]{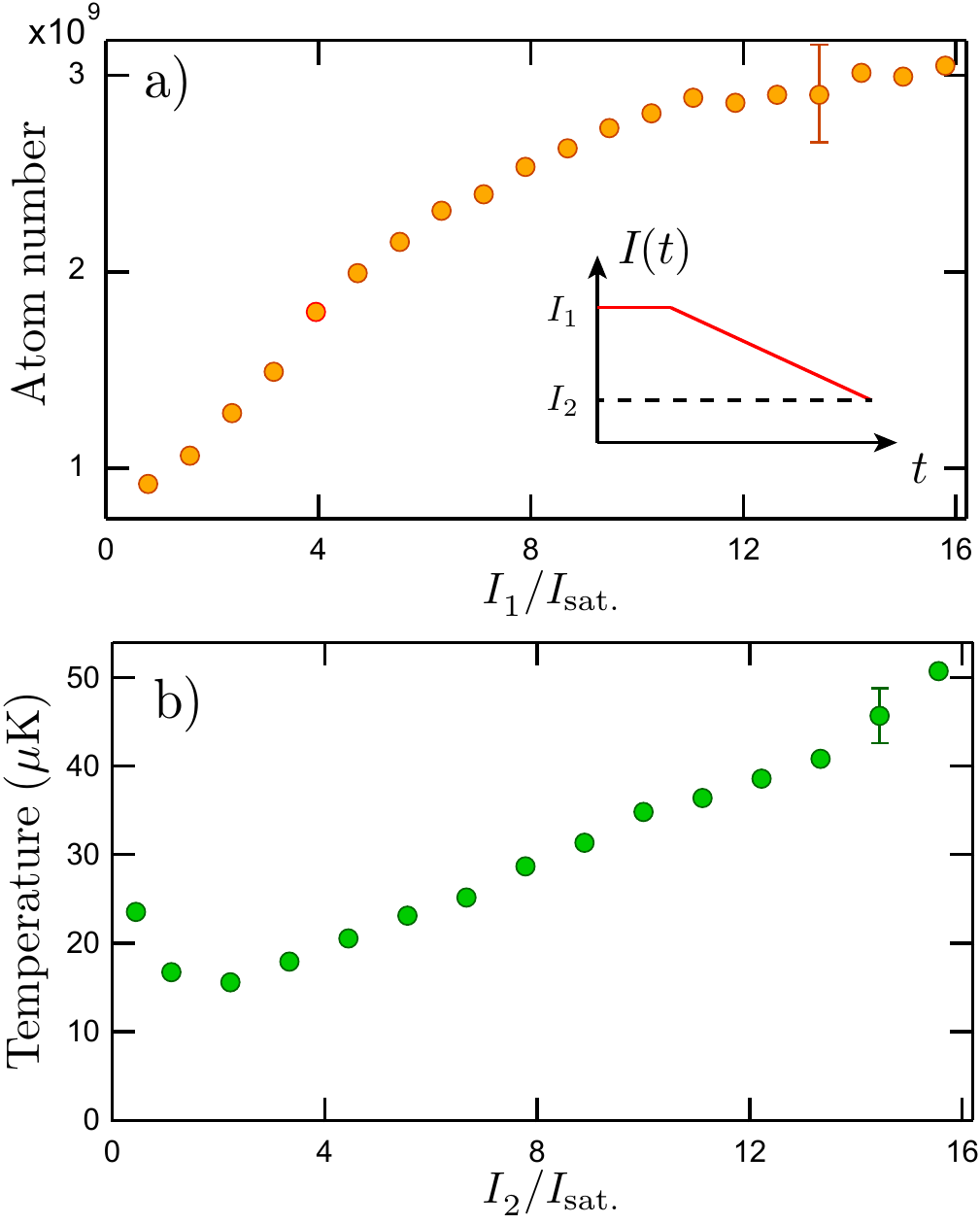}
	\caption{a) Number of atoms in the gray molasses, as a function of the total initial intensity $I_1$ of the D1 beams. b) Temperature of the gray molasses as a function of the light intensity $I_2$ at the end of the power ramp.}
	\label{fig:Fig3SM}
\end{figure}

The molasses sequence is performed in two steps. As mentioned above, the capture efficiency of the atoms into the gray molasses depends crucially on the initial temperature of the cloud (see Fig.\ref{fig:Fig2SM}.b). To maximize the capture efficiency, we first turn on the D1 laser and its power is maintained constant for 2 ms, at maximum value. This stage favors capturing a maximum number of atoms from the CMOT into the gray molasses. This can be seen in Fig.\ref{fig:Fig3SM}.a), which shows the evolution of the atom number captured in the gray molasses as a function of the total intensity of D1 beams. This number increases linearly at low light intensities, and starts to inflect above a certain value ($\simeq 6 I_{\rm sat}$). Our data shows that the saturation is almost reached at the maximum available D1 power, corresponding to $\simeq 16 I_{\rm sat}$. At this stage, the temperature of the cloud is $50\ \mu$K. 

Subsequently, we linearly decrease the D1 power during 10 ms, which helps to further lower the temperature of the cloud. Fig.\ref{fig:Fig3SM},b) shows the dependence of the gray molasses temperature as a function of the final intensity value at the end of the ramp. We observe an optimal value of the total intensity around $2.1\times I_{\rm sat}$ per beam, which corresponds to a temperature of $16 \pm 3 \mu K$. No size increase of the atomic cloud can be observed during the gray molasses stage, and the density remains roughly equal to that obtained at the end of the CMOT.

To summarize, at the end of the gray molasses stage we obtain $3.10^{9}$ atoms at a density of $2.2\times 10^{11}$~cm$^{-3}$ and temperature of $16\ \mu$K. This corresponds to a phase space density of $7\times 10^{-5}$ (i.e. a $10^4$ increase compared to the initial D2 MOT stage) and represents a good starting point towards further evaporative cooling towards Bose-Einstein condensation of $^{41}$K atoms~\cite{Chen:PRA:16}.


%